\documentclass[twocolumn,preprintnumbers,superscriptaddress,nofootinbib,aps,prd,floatfix]{revtex4-2}

\usepackage{enumerate}
\usepackage{amsmath,amssymb}
\usepackage{graphicx}
\usepackage{slashed}
\usepackage{xspace,slashed}
\usepackage{hyperref}
\hypersetup{colorlinks=true, citecolor=blue, urlcolor=blue, linkcolor=blue}
\usepackage[normalem]{ulem}
\usepackage{subfigure,orcidlink}
\usepackage{multirow,array}

\usepackage{hyperref}
\hypersetup{colorlinks=true, citecolor=blue, urlcolor=blue, linkcolor=blue}

\newcommand{\be}{\begin{eqnarray*}}
\newcommand{\ee}{\end{eqnarray*}}

\newcommand{\bee}{\begin{eqnarray}}
\newcommand{\eee}{\end{eqnarray}}
\newcommand{\beeq}{\begin{equation}}
\newcommand{\eeeq}{\end{equation}}

\newcommand{\beq}{\begin{eqnarray}} 
\newcommand{\eeq}{\end{eqnarray}}

\begin{document}

\title{Double and Triple Higgs Production to probe the Electroweak Phase Transition}

\begin{abstract}
The production of three Higgs bosons could be a stretch goal for the LHC and a strategic case for future colliders. In this work, we analyse the phenomenological prospects of (neutral) triple Higgs compared to di-Higgs boson production, for a range of Higgs-sector extensions from a strong first-order electroweak phase transition perspective. In parallel, we include constraints from existing exotics and Higgs boson measurements that further limit the parameter space of such models. Resonance contributions offer large modifications in particular for triple Higgs production, albeit starting from a small SM expectation. With enhancements of order 40 over the SM, however, experimental efforts to obtain limits at the HL-LHC are well-motivated and well-placed. This is further highlighted by the potential of these processes to inform the investigation of the thermal history of our universe. 
\end{abstract}

\author{Lisa~Biermann\orcidlink{0000-0003-1469-6400}}\email{lisa.biermann@kit.edu}
\affiliation{Institute for Theoretical Physics, Karlsruhe Institute of Technology, 76128 Karlsruhe, Germany\\[0.1cm]}
\author{Christoph~Borschensky\orcidlink{0000-0002-8486-8784}}\email{christoph.borschensky@kit.edu}
\affiliation{Institute for Theoretical Physics, Karlsruhe Institute of Technology, 76128 Karlsruhe, Germany\\[0.1cm]}
\author{Christoph~Englert\orcidlink{0000-0003-2201-0667}}\email{christoph.englert@glasgow.ac.uk}
\affiliation{School of Physics and Astronomy, University of Glasgow, Glasgow G12 8QQ, U.K.\\[0.1cm]}
\author{Margarete~M\"uhlleitner\orcidlink{0000-0003-3922-0281}}\email{milada.muehlleitner@kit.edu}
\affiliation{Institute for Theoretical Physics, Karlsruhe Institute of Technology, 76128 Karlsruhe, Germany\\[0.1cm]}
\author{Wrishik~Naskar\orcidlink{0000-0002-4357-8991}}\email{w.naskar.1@research.gla.ac.uk}
\affiliation{School of Physics and Astronomy, University of Glasgow, Glasgow G12 8QQ, U.K.\\[0.1cm]}

\maketitle

\preprint{KA-TP-16-2024}

\section{Introduction}
\label{sec:intro}
As the Higgs boson programme at the Large Hadron Collider (LHC) is entering a period of intense data gathering, rare processes become experimentally accessible and, hence, relevant. A particularly motivated process in this regard is the production of multiple Higgs bosons; di-Higgs boson production and associated analyses are cornerstones of the Beyond the Standard Model (BSM) physics programme at the LHC and beyond. Searches for Higgs pair production in gluon fusion $gg\to hh$ and weak boson fusion $pp \to hh jj$ are fairly well-established experimentally, cf.~e.g.~\cite{ATLAS:2023qzf,CMS:2022hgz}. With that in mind, both the theory and the experimental communities have started to increasingly look towards sensitivity to triple Higgs production. Perhaps a stretch goal in the SM context for the LHC even during its high-luminosity (HL) phase, triple Higgs production could indeed be a strategic target for future colliders~\hbox{\cite{Chen:2015gva,Fuks:2015hna,Papaefstathiou:2015paa,Fuks:2017zkg,Abdughani:2020xfo,Papaefstathiou:2019ofh,Papaefstathiou:2020lyp,Kilian:2017nio,Papaefstathiou:2023uum,Lane:2024vur}}. Most targeted search strategies relying on a variety of optimisation methods and development, cf.~e.g.~\cite{Stylianou:2023xit,Belvedere:2024wzg}, will be the key to background mitigation and signal isolation of these final states.

To what extent does triple Higgs production add information to the presently established Higgs programme? The LHC should be able to become sensitive to SM Higgs pair production during the HL-LHC phase~\cite{Cepeda:2019klc}. In parallel, a fine-grained picture of a {\emph{single}} Higgs boson's interaction with other known matter should be well-established at that point~\cite{Cepeda:2019klc,deBlas:2019rxi}. If we tension an SM-resembling outcome of the LHC at this point with, e.g., the leading Standard Model version of effective field theory (SMEFT), triple Higgs production will not add a significant amount of BSM-discriminative power to a top-level analysis of LHC data once single and double Higgs production are included. This results from the correlation of different Higgs multiplicities due to ${SU}(2)_{L}$ gauge invariance~\cite{Grzadkowski:2010es}. Manoeuvring away from the SMEFT assumptions, a potential non-linear character of the electroweak scale, chiefly expressed through Higgs Effective Field Theory (HEFT), treats different Higgs multiplicities as independent parameters and thus adds ad-hoc independence of different Higgs interactions~\cite{Herrero:2021iqt,Herrero:2022krh,Delgado:2023ynh}. In this case, we can expect a non-resonant enhancement of the triple Higgs rate of about an order of magnitude over the SM expectation of ${\cal{O}}(50~\text{ab})$ at the LHC~\cite{deFlorian:2019app}, which would leave no discernible imprint in single or double Higgs physics~\cite{Anisha:2024ljc}.

The appearance of resonantly produced states can alter these EFT expectations and lead to very large enhancements of multi-Higgs rates~\cite{Basler:2018dac,Stylianou:2023xit,Abouabid:2021yvw} (see also~\cite{Dawson:2023ebe} for a recent discussion on the relation of the two-Higgs doublet model (2HDM) with EFT). Whilst the LHC is closing in on the production of Higgs and top-philic SM extensions, there is considerable space left for such scenarios to evade current constraints~\cite{Basler:2019nas}. This holds in particular for non-minimal extensions of motivated theories such as the two-Higgs doublet models. Consequently, there is good motivation for going beyond Higgs pair production in such cases, potentially with LHC-relevant implications. 

Adding to this, the Higgs potential that is fingerprinted by the production of different Higgs multiplicities plays a fundamental role in theories that connect baryogenesis with the electroweak scale. The electroweak cross-over of the SM does not adequately address Sakharov's criteria~\cite{Sakharov:1967dj}, and any modification of the potential to address this shortfall is expected to leave phenomenological imprints on multi-Higgs production related to a strong first-order electroweak phase transition (SFOEWPT).

In this note, we take these observations as motivation to analyse the interplay of multi-Higgs final states at colliders from the angle of an SFOEWPT (see also~\hbox{\cite{Li:2019tfd,Ramsey-Musolf:2019lsf,Huang:2017jws,No:2013wsa,Basler:2019nas}}). To this end, we consider a range of Higgs sector extensions and elaborate on the expected modifications of the triple Higgs rate at the LHC and future experiments. We compare our findings with similar considerations for Higgs pair production to not only gauge the level of complementarity that $hh$ and $hhh$ production can provide but also to analyse the critical range of enhancements of the $hh(h)$ rates that can be achieved from the viewpoint of these scenarios and an SFOEWPT. As we will see, the phenomenological relation of multi-Higgs cross section enhancements and an SFOEWPT is not direct, but significant $\sim 40$ enhancements for scenarios with an SFOEWPT can be obtained in the $hhh$ channel with notable complementarity compared to $hh$ final states. This adds to the motivation for extending the multi-Higgs programme at the LHC to triple Higgs production.

This work is organised as follows: In Sec.~\ref{sec:models} we quickly review the (extended) two-Higgs doublet scenarios on which we will base our analysis to make this work self-contained. There, we also provide details of our scan and event simulation methodology. Section~\ref{sec:results} is devoted to a detailed discussion and comparison of the $hh$ and $hhh$ rates, and their potential relation with an  SFOEWPT. We conclude in Sec.~\ref{sec:conc}.

\section{Models, Scans and Cross Sections}
\label{sec:models}
We focus on resonant extensions of the SM which have the potential to enhance cross sections through new on-shell contributions that are not present in the continuum production expected in the SM. Modifications of inter-Higgs couplings ubiquitous in Higgs sector extensions with related interference effects between different new physics and modified SM contributions can lead to an intricate interplay of different phase space regions. Together these can lead to either a net enhancement or a reduction of the cross sections of $hh(h)$ production. Whilst some differential information might be obtainable for the $hh$ final states at the HL-LHC, this is less likely for $hhh$ production. However, experimental proof-of-principle investigations have yet to be made available. We will therefore focus on the integrated cross sections as relevant physical observables but will touch on the interplay of different phase space regions that highlight resonance and interference contributions in relation to the SM.

Along these lines, a particularly motivated class of models is the extension of the SM by an additional doublet in 2HDMs and the latter's complex and singlet-extended variation. These models introduce all aforementioned phenomenological modifications in clear relation to existing Higgs measurements and exotics searches; they further enable a direct correlation of multi-Higgs rates with an SFOEWPT (see in particular~\cite{Karkout:2024ojx}). This is most transparently expressed through the so-called ``real'' 2HDM (R2HDM) where the 2HDM parameters are assumed to be real variables thus making clear distinctions between CP-even and -odd exotic Higgs bosons. The R2HDM is generalised by admitting complex phases (giving rise to the ``complex" 2HDM, C2HDM) which provides a natural interface to incorporate CP-violation as an avenue to satisfy Sakharov's criteria further. Extensions of the R2HDM by an additional real singlet field (the Next-to-Minimal 2HDM, N2HDM) generalise the typical correlations of the R2HDM through additional contributions to the potential and a richer mixing structure~\cite{Muhlleitner:2016mzt} of states with given CP quantum numbers. In the following, we will consider these three scenarios to discuss correlations and their modifications when requiring a sufficient SFOEWPT. We now swiftly review these models to make this work self-contained.

\subsubsection*{Variations on the 2HDM}
The potential for the 2HDM is given by
\begin{equation}
\label{eq:r2dm}
\begin{split}
V_{\text{2HDM}}&=m_{11}^{2}\Phi_{1}^{\dagger}\Phi_{1}+m_{22}^{2}\Phi_{2}^{\dagger}\Phi_{2} - \left( m_{12}^{2}\Phi_{1}^{\dagger}\Phi_{2}+{\text{h.c.}} \right)\\
&+\frac{\lambda_{1}}{2}\left(\Phi_{1}^{\dagger}\Phi_{1}\right)^{2}+\frac{\lambda_{2}}{2}\left(\Phi_{2}^{\dagger}\Phi_{2}\right)^{2}\\
&+\lambda_{3}\left(\Phi_{1}^{\dagger}\Phi_{1}\right)\left(\Phi_{2}^{\dagger}\Phi_{2}\right)+\lambda_{4}\left(\Phi_{1}^{\dagger}\Phi_{2}\right)\left(\Phi_{2}^{\dagger}\Phi_{1}\right)\\
& + \left[\frac{\lambda_{5}}{2}\left(\Phi_{1}^{\dagger}\Phi_{2}\right)^{2}+\text{h.c.}\right],
\end{split}
\end{equation}
with 
\begin{equation}
\Phi_{i}={1\over \sqrt{2}} \left( \begin{matrix} \phi_i^+ \\ \varphi_i + i a_i \end{matrix} \right),~i=1,2,
\end{equation}
transforming as  $({\bf{1}},{\bf{2}})_{1/2}$ under the SM gauge group $SU(3)_C\times SU(2)_L\times U(1)_Y$. The potential obeys the usual ${\mathbb{Z}}_2$ symmetry assignments to remove flavour-changing interactions~\cite{Glashow:1976nt}, which is softly broken by the $m_{12}^2$ term in $V_{\text{2HDM}}$. In the case of the R2HDM, all parameters in this potential are taken to be real 
\begin{equation}
 m_{11}^2,m_{22}^2,m_{12}^2,\lambda_{1,\dots,5} \in {\mathbb{R}}\quad\hspace{0.9cm} \text{(R2DHM)},
\end{equation}
whereas in the CP-violating version 
\begin{equation}
 m_{11}^2,m_{22}^2,\lambda_{1,\dots,4} \in {\mathbb{R}},~
 m_{12}^2, \lambda_5 \in {\mathbb{C}}\quad \text{(C2DHM)}
\end{equation}
(with independent phases for $m_{12}^2$ and $\lambda_5$ so that they cannot be absorbed by field redefinitions). 

The usual equations to obtain the minimum for this potential alongside the physical masses and mixing angles hold, e.g., for the R2HDM the physical neutral Higgs masses are related to the Lagrangian eigenstates via
\begin{equation}
\left(
\begin{matrix}
H\\
h
\end{matrix}
\right) = R_2(\alpha)
\left(
\begin{matrix}
\varphi_1\\
\varphi_2
\end{matrix}
\right),
\end{equation}
where $R_2(\alpha)$ is the standard 2-dimensional orthogonal rotation matrix. Furthermore,
\begin{equation}
\tan \beta = {\langle \Phi_2 \rangle \over \langle \Phi_1\rangle} = {v_2\over v_1} 
\end{equation}
denotes the diagonalisation angle of $R_2(\beta)$ for the CP-odd and charged fields (including the massless Goldstone mode). The brackets around the fields denote their respective vacuum expectation values (VEVs), with $\sqrt{v_1^2+v_2^2} \equiv v \approx 246.22$~GeV.
  
In the C2HDM, due to explicit CP violation, one can still define massless Goldstone modes through $R_2(\beta)$ (i.e. the vacuum expectation values $\langle \Phi_{1,2}\rangle$ can be aligned), but the neutral electromagnetic fields no longer have defined CP quantum numbers and are now diagonalised by a 3-dimensional orthogonal rotation matrix $R_3(\alpha_1,\alpha_2,\alpha_3)$. 

Extending the 2HDM with an additional real singlet under the SM gauge group $\Phi_S\sim ({\bf{1}},{\bf{1}})_0$ leads to the N2HDM
\begin{equation}
\label{eq:n2dm}
\begin{split}
V_{\text{N2HDM}}&=V_{\text{R2HDM}}+\frac{1}{2}m_{S}^{2}\Phi_{S}^{2}+\frac{\lambda_{6}}{8}\Phi_{S}^{4}\\&+\frac{\lambda_{7}}{2}\left(\Phi_{1}^{\dagger}\Phi_{1}\right)\Phi_{S}^{2}+\frac{\lambda_{8}}{2}\left(\Phi_{2}^{\dagger}\Phi_{2}\right)\Phi_{S}^{2}\,,
\end{split}
\end{equation}
which generalises the mixing in the CP-even sector compared to the R2HDM to a 3-dimensional orthogonal rotation after expanding $\Phi_S$ around a non-vanishing VEV $\langle \Phi_S \rangle$, $\Phi_S=\langle \Phi_S \rangle + \varphi_3$. The potential is invariant under the usual (softly broken) $\mathbb{Z}_2$ symmetry and an additional $\mathbb{Z}_2^\prime$ symmetry, under which $\Phi_{1,2} \to \Phi_{1,2}$ and $\Phi_S \to - \Phi_S$.

In all these scenarios the interactions of the Higgs boson(s) with known matter are changed as a consequence of the ${\mathbb{Z}}_2$ assignments giving rise to the usual classification of 2HDM models~(see~Ref.~\cite{Branco:2011iw} for a review) as well as mixing. We will focus on the type-I scenario in the following as this provides a wider range of acceptably strong phase transitions~\cite{Basler:2016obg,Atkinson:2021eox,Atkinson:2022pcn}. The couplings of the extended Higgs spectrum to the SM quarks in the R2HDM are then given by 
\begin{equation}
\begin{split}
\label{eq:yuk1}
\xi_h^{u,d} &= {\cos\alpha\over \sin\beta} \,, \\ 
\xi_H^{u,d} &= {\sin\alpha \over \sin\beta} \,, \\ 
\xi_A^{u,d} & = \cot\beta\,, \\ 
\end{split}
\end{equation}
relative to the SM Higgs couplings. In the extended 2HDM, these couplings are then further modified by the aforementioned rotations in the C2HDM and N2HDM, respectively.

The inter-Higgs interactions obtained from expanding the potential in the mass basis further impact the multi-Higgs production through modified trilinear and quartic Higgs boson interactions. Depending on the realisations we consider, these can be very different, ranging from relatively rigid correlations in the 2HDM to relaxing them in the N2HDM and the C2HDM along the lines of singlet admixture and explicit CP violation, respectively. This will enable us to comment on the impact of these various phenomenological variations on the expected $hh(h)$ production rates in light of the strength of the electroweak phase transition. A notable exception to the phenomenology we discuss is provided by the 2HDM with additional symmetry protection that achieves alignment through an enhanced custodial symmetry~\cite{Battye:2011jj,Pilaftsis:2011ed,Dev:2015bta}.

\subsubsection*{Scanning for SFOEWPTs and multi-Higgs cross sections}
Throughout this work, we will consider a Higgs mass of
\begin{equation}
m_h=125.09~\text{GeV}
\end{equation}
and use {\tt{ScannerS}}~\cite{Coimbra:2013qq,Muhlleitner:2020wwk} for the generation of viable parameter points, where all other input parameters (such as the exotic Higgs masses and mixing angles) are varied randomly in wide ranges to cover the parameter space allowed by the relevant theoretical constraints and by the experimental constraints given by the Higgs measurements (as checked via an interface to {\tt{HiggsTools}}~\cite{Bechtle:2020pkv,Bechtle:2020uwn,Bahl:2022igd}), flavour constraints~($\mathcal{R}_b$~\cite{Haber:1999zh,Deschamps:2009rh} and $B\rightarrow X_s \gamma $~\cite{Deschamps:2009rh,Mahmoudi:2009zx,Hermann:2012fc,Misiak:2015xwa,Misiak:2017bgg, Misiak:2020vlo}), and electroweak precision data (by demanding the $S$, $T$, and $U$ parameters \cite{Peskin:1991sw} to be within 2$\sigma$ of the SM fit \cite{Baak:2014ora}).

We analyse the finite temperature phenomenology using the recently released version 3 of {\tt{BSMPT}}~\cite{Basler:2024aaf,Basler:2020nrq,Basler:2018cwe}. We will focus on the strength of the phase transition during the so-called percolation stage and we deem a phase transition to be of first order and strong when
\begin{equation}
\xi_p = {v_p(T_p)\over T_p} > 1\,,
\end{equation}
i.e.\ when the VEV at the percolation temperature $T_p$ is larger than unity in units of $T_p$. This is a conventional measure to safeguard against sphaleron washout effects when considering electroweak baryogenesis (see also~\cite{Espinosa:2015qea,Patel:2011th} for further discussions). For a given parameter choice that is consistent with the theoretical constraints and the current experimental outcome as verified by {\tt{ScannerS}}, we trace the thermal history of the corresponding model. {\tt{BSMPT}} achieves this by including the temperature-independent potential up to one-loop order
\begin{multline}
V_{\text{eff}}(T) = V_0(T=0) + V_{\text{CW}}(T=0) + V_{\text{CT}}(T=0) \\ + V_{\text{T}}(T) + V_{\text{daisy}}(T)
\end{multline}
alongside thermal corrections~(see e.g.~\cite{Quiros:1994dr,Quiros:1999jp,Morrissey:2012db} for excellent reviews of the subject) as well as Daisy-resummation (concretely we consider these in the Arnold-Espinosa approach~\cite{Arnold:1992rz}). As done in Refs.~\cite{Basler:2016obg,Basler:2019iuu,Basler:2021kgq}, we include finite counterterm contributions to identify the minima of the potential, the Higgs masses and mixing angles at tree level ($V_0$) and at one loop ($V_0 + V_{\text{CW}} + V_{\text{CT}}$) at $T=0$, to facilitate tests for the compatibility with experimental constraints. Here $V_{\text{CW}}$ and $V_{\text{CT}}$ denote the one-loop Coleman-Weinberg and the counterterm potential, respectively.
With {\tt BSMPT} we furthermore compute the trilinear and quartic Higgs self-couplings from the third and fourth derivatives of the potential, respectively, which are used, at tree level, as an input to the $hh(h)$ cross sections.

To compare the results of the electroweak phase transition with the expected LHC phenomenology, we implement the various (extended) 2HDMs using {\tt{FeynRules}}~\cite{Alloul:2013bka} and {\tt{NLOCT}}~\cite{Degrande:2014vpa} which is interfaced via {\tt{Ufo}}~\cite{Degrande:2011ua,Darme:2023jdn} with {\tt{MadGraph\_aMC@NLO}}~\cite{Alwall:2014hca}. Cross sections are generated for $hh(h)$ production through gluon fusion, $pp\to hh(h)$, at leading, i.e.~one-loop order, including top and bottom quark contributions (derived from Eq.~\eqref{eq:yuk1} directly or as part of the parameter scan). Points from the {\tt ScannerS+HiggsTools+BSMPT} scan are interfaced with this toolchain for an LHC centre-of-mass energy $\sqrt{s} = 13~\text{TeV}$ to compare $\xi_p$ and cross section modifications of multi-Higgs modes at the LHC. It is known that the dominant QCD corrections that largely increase the parton-level leading order cross section generalise qualitatively to BSM resonance structures, see e.g.~\cite{Djouadi:1991tka}.

\begin{figure}[!t]
  \centering
  \includegraphics[width=0.38\textwidth]{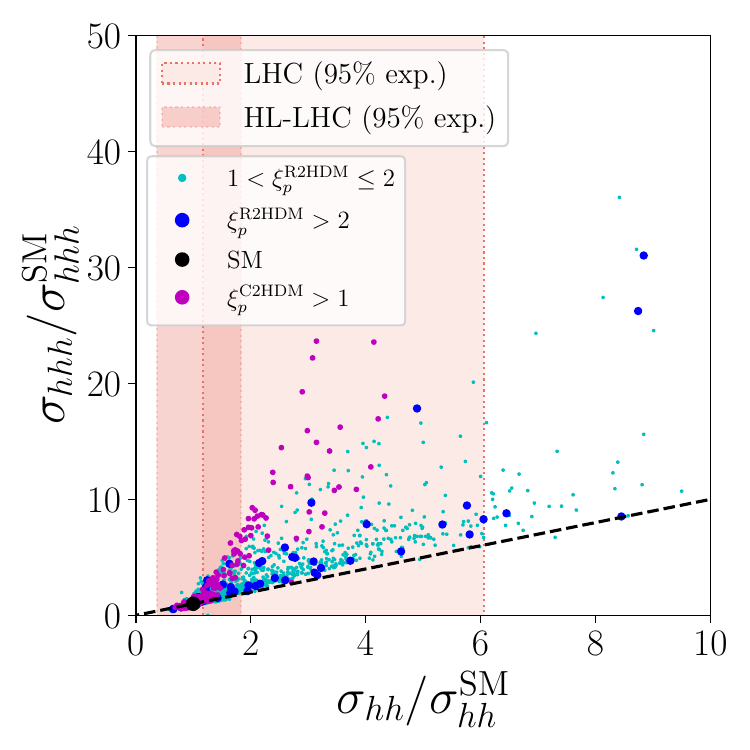}
  \caption{\label{fig:r2hdmscan} 2HDM scan results for gluon fusion $hh(h)$ production relative to the non-resonant SM expectation for $\xi_p > 1$ at the LHC and $13~\text{TeV}$ collisions. The colors denote the various phase transition strengths in the R2HDM and the C2HDM. The red shaded region shows the current and HL-LHC $hh$ sensitivity \cite{CMS:2022dwd}. \label{ref:r2hdm_scans}}
  \vspace{-0.5cm}
\end{figure}

\begin{figure*}[!t]
  \centering
  \includegraphics[width=0.43\linewidth]{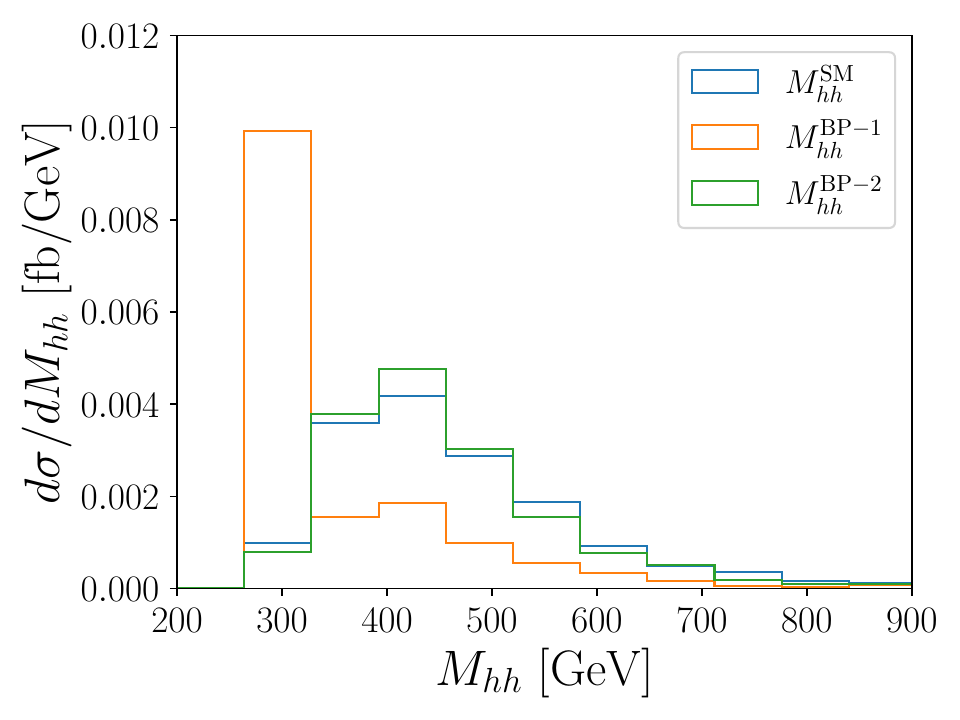}\hspace{1cm}
  \includegraphics[width=0.43\linewidth]{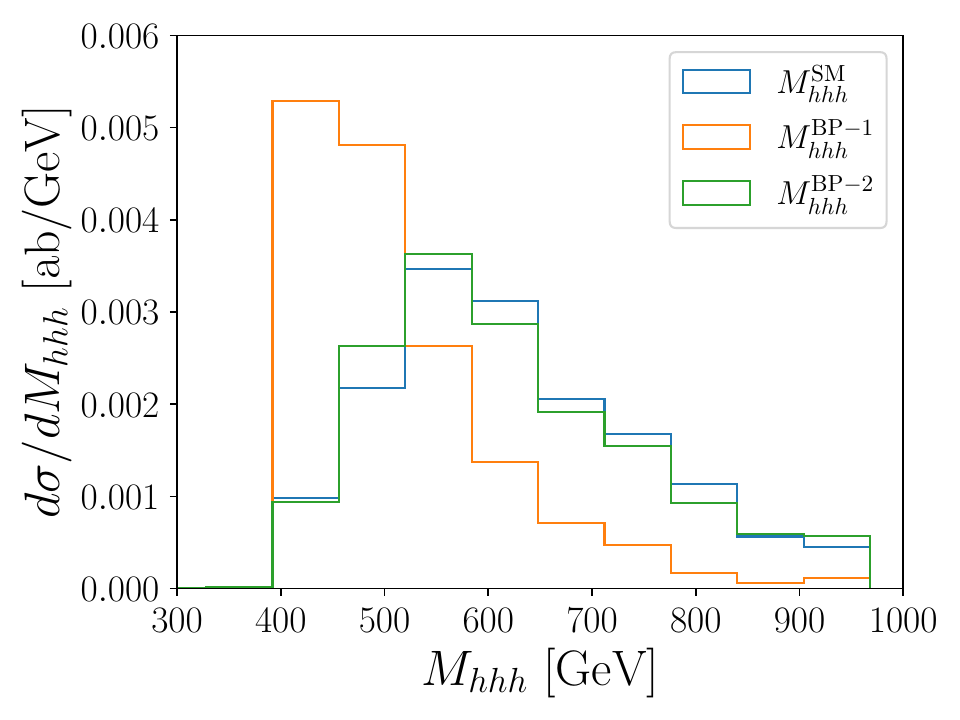}
  \caption{\label{fig:minv} Invariant di- and triple- Higgs mass distributions for the SM as well as for 
  benchmark points with a large (BP-1) and representatively small (BP-2) enhancement for the $hhh$-production cross sections. The respective cross-sections and parameter choices for the points are shown in Tab.~\ref{tab:minmax}. }
\vspace{-0.2cm}
\end{figure*}

\begin{table*}[ht]
\centering
\begin{tabular}{|l||c|c|c|c|c|c|c|c|c|c|}
\hline
Benchmark Points & $\sigma_{hh}/\sigma^{\mathrm{SM}}_{hh}$ & $\sigma_{hhh}/\sigma_{hhh}^{\mathrm{SM}}$ & $M_H~[\text{GeV}]$ & $\Gamma_H~[\text{GeV}]$& $C_{ffh}$ & $C_{ffH}$  & $g_{hhh}~[\text{GeV}]$ & $g_{hhH}~[\text{GeV}]$ & $g_{hhhh}$ & $g_{hhhH}$ \\ \hline \hline
largely enhanced & 3.24 & 15.26 & 274.29 & 0.20 & 1.027 & -0.124 & 167.26 & 75.28 & 0.661 & 0.203 \\ \hline
SM-like & 1.02 & 1.02 & 469.30 & 2.49 & 0.997 & -0.484 & 190.54 & -7.11 & 0.774 & -0.011 \\ \hline
\end{tabular}
\caption{Cross sections and parameter choices for chosen benchmark points with large (BP-1) and representatively small (BP-2) enhancements in $hhh$-production cross sections.}
\label{tab:minmax}
\vspace{-0.2cm}
\end{table*}

\section{Results and Discussion}
\label{sec:results}
We now turn to results. As all scenarios are characterised by fixed charge assignments for the electroweak group, we can expect the $hh$ and $hhh$ phenomenology to be tightly correlated. Hence, anomalies should statistically show up first in the more abundant $hh$ channels. Any resonance structure, however, can be a priori relevant to the extent that the continuum SM expectation for 
\begin{equation}
\label{eq:hhhxsec}
\sigma(pp\to hhh)\simeq 50~\text{ab}
\end{equation} 
at the LHC~\cite{deFlorian:2019app} is highly misleading. For comparison, the gluon fusion Higgs pair production cross section is at next-to-leading order QCD around~\cite{Plehn:1996wb,Dawson:1998py,Borowka:2016ypz,Baglio:2020ini}  
\begin{equation}
\sigma(pp\to hh)\simeq 27~\text{fb}
\end{equation}
in the SM for 13 TeV collisions. In such an instance, resonances could raise the cross section to a level where the $hhh$ efforts at the HL-LHC could indeed provide a relevant cross-check of a potential discovery at the LHC in the $hh$ channels. Depending on the scenario, however, it is also possible that deviations could manifest themselves through large enhancements of $hhh$ production exclusively. Either of these observations could then further pave the way for a subsequent analysis at a future hadron facility such as the FCC-hh. We will discuss the implications for such a future collider in passing.

In Fig.~\ref{ref:r2hdm_scans}, we show the obtained cross sections for Higgs pair and triple Higgs production, normalised to the respective SM expectation for the R2HDM and the C2HDM scan for phase transition strengths $\xi_p>1$. On the one hand, underproduction relative to the SM is possible, but phenomenologically largely uninteresting. On the other hand, new contributions can greatly enhance the $hh(h)$ rates. As expected, the $hh$ and $hhh$ cross sections show a clear correlation, yet the relative enhancement tends to be greater for $hhh$ production. A fully differential analysis of triple Higgs final states will be a serious experimental challenge, yet the 
\begin{equation}
M^2_{hhh}=(p_{h_1}+p_{h_2}+p_{h_3})^2
\end{equation} 
distribution is a particularly telling observable to quantitatively understand the observed cross section modifications, see e.g. Fig.~\ref{fig:minv}. 

In case the heavy state is close to the $m_H\simeq 2m_h$ threshold, $pp\to hh$ receives a large enhancement in the resonance region which in turn is accompanied by underproduction for $M^2_{hh}=(p_{h_1}+{p_{h_2}})^2\gtrsim m_H^2$ compared to the SM as these contributions effectively enhance the destructive interference between the involved triangle and box topologies~\cite{Dolan:2012rv,Arco:2022lai,Heinemeyer:2024hxa}. This situation is mirrored for $pp\to hhh$ where the threshold region $M_{hhh}\simeq 3m_h$ probes a wider range of $M_{{h_i}{h_j}}\simeq 2m_H$, $i,j=1,2,3,~j> i$ enhancement (effectively isolating $pp\to Hh, H\to hh$). This can lead to a comparably larger enhancement of $hhh$ production compared to $hh$. In our scan, we can find such maximum enhancements up to factors of $\sim 4$ of $hhh$ over $hh$ production (relative to their respective SM values) for the R2HDM.

For parameter choices that lead to heavier spectra, enhancements can again arise from $H\to hh$ resonance structures. These are, however, small when compared to enhancements arising from three body decays $H\to hhh$. While the latter correspond to small branching ratios of the typically top-philic states, the excess over the relatively small SM expectation can be considerable.

\begin{figure}[!b]
  \centering
  \includegraphics[width=0.9\linewidth]{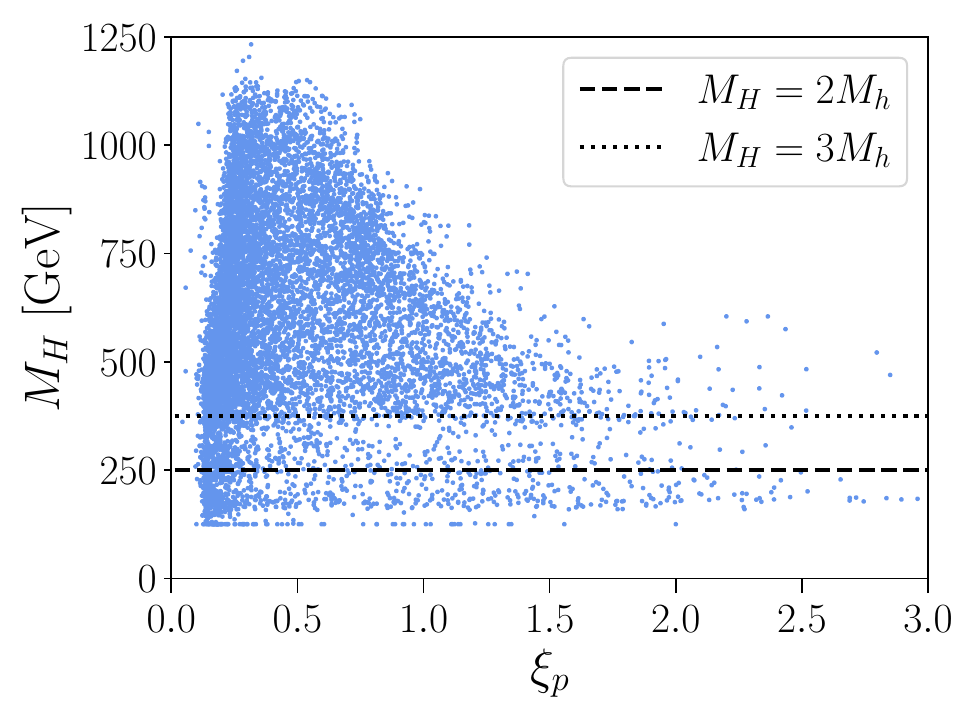}
  \caption{\label{fig:masssp} Mass spectra, here specifically for the resonance $H$, explored in the scan. For details see text.}
\end{figure}

How do these observations relate to the strength of the first-order phase transition? As a thermodynamic process, the EWPT is driven by the physics of the light degrees of freedom. For the concrete example of the R2HDM with relatively rigid coupling constraints of the Higgs bosons to other matter (in particular when considering the Higgs signal strength constraints inferred from LHC measurements), this is achieved by making the non-SM degrees of freedom more accessible via lighter spectra. This is visible from Fig.~\ref{fig:masssp}, which distils the results of a scan of the strength of the phase transition $\xi_p$ at the percolation stage as a function of the mass of the neutral CP-even heavy Higgs boson $H$. 

A priori this is good news for multi-Higgs final states and their observation at the LHC. Triple Higgs production can receive new contributions from nested $Hh$ production but, more importantly, new resonant $H\to hhh$ decays. Cross section enhancements for moderately strong transitions\footnote{In fact, we find only percent-level differences between the nucleation and percolation temperatures. For the models studied in this work, the distinction between these temperatures is somewhat irrelevant, however, they have very different meanings for the phenomenology of the early universe, see~\cite{Ellis:2018mja,Athron:2023rfq}.} $ \xi_p\gtrsim1$, are $\sigma_{hhh}/\sigma^\text{SM}_{hhh}\simeq 40 $ and $\sigma_{hh}/\sigma^\text{SM}_{hh}\simeq 20$ (we will comment on the relation with an SFOEWPT below as these numbers can also be found for $\xi_p<1$.). Enhancements of this size in the triple Higgs rate of Eq.~\eqref{eq:hhhxsec} amount to around 140 events in the $6~b$ final state after applying a 70\% b-tagging efficiency at the HL-LHC phase ($13~\text{TeV}$, $3~\text{ab}^{-1}$). Whilst this is undeniably challenging there the appearance of additional resonance structures in the b jet distribution can be exploited to combat backgrounds. Cross-relating such an enhancement with $hh$ and exotics measurements will enable the phenomenological dissection of the TeV scale's relation with electroweak baryogenesis: To guide the eye in identifying which enhancements can be relevant from $hh$ limits, the red bands in Fig.~\ref{fig:r2hdmscan} highlight estimates of current and HL-LHC $hh$ sensitivity~\cite{CMS:2022dwd}. 

\begin{figure}[!t]
  \centering
  \includegraphics[width=0.38\textwidth]{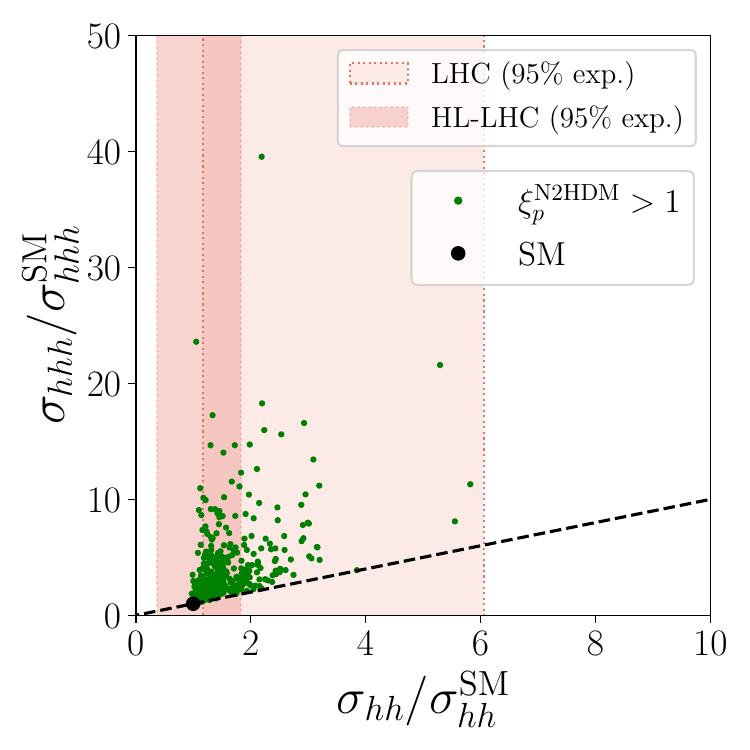}
  \caption{Double and triple Higgs production cross sections relative to the SM expectation for a scan over the N2HDM for $\xi_p > 1$ at the LHC (for a centre of mass energy of $13~\text{TeV}$).
  \label{ref:n2hdm_scans} 
  }
  \vspace{-0.5cm}
\end{figure}

Moving to stronger phase transitions, the three-Higgs threshold $H\to hhh$ becomes decreasingly relevant and most observations are from correlated resonant $H\to hh$ and $Hh\to hhh$ contributions. There is a wider phase space available for $Hh\to hhh$ compared to $H\to hh$ which leads to a comparably larger enhancement in $hhh$ production, however, compared to a small baseline rate expected in the SM. As the $hh$ rate starts from a much larger expectation compared to $hhh$ production, it is clear that the former will statistically dominate the multi-Higgs phenomenology at the HL-LHC if agreement with the SM prevails. Should a discovery be made, factors of ten enhancements in the triple Higgs rate might provide further motivation to target $hhh$ production at the HL-LHC. The multi-resonant structures described above will then also contribute to further control backgrounds compared to non-resonant considerations, which are harder to isolate from continuum backgrounds. Overall, the strength of the phase transition is predominantly driven by the mass scale of the non-SM states. These will leave correlated effects in the rates, but the neutral Higgs rates alone are not indicative of the strongness of the phase transition in the R2HDM.

Adding additional resonances to the spectrum, however, opens up significant parameter space for non-standard phenomenology. This can be observed in the cross-sections in the complex generalisation of the R2HDM, the C2HDM, along $hhh$ compared to $hh$, which is overlayed in Fig.~\ref{ref:r2hdm_scans}. The cross section enhancements discussed above, i.e. $H\to hh$ and $H\to hhh$, can appear simultaneously when the two additional scalar degrees of freedom in the C2HDM are kinematically accessible in two- and three-body decays, respectively. The phenomenology of the $hh(h)$ final states is more involved. Due to the stringent limits from the measurements of the electric dipole moment, however, \cite{Roussy:2022cmp} (cf.~the discussion of the impact on the C2HDM in \cite{Fontes:2017zfn,Biekotter:2024ykp}) complex phases of the top- and bottom-Higgs interactions are quantitatively small as are the Higgs mixings that are absent in the R2HDM, so that no dramatic departure from the R2HDM paradigm is observed (or expected). This can also be visualised looking at the amount of CP admixture, which is a measure of the SM-like scalar-pseudoscalar mixing. For the Type-I C2HDM, following Refs.~\cite{Muhlleitner:2017dkd,Fontes:2017zfn}, the pseudoscalar admixture is defined as,
\begin{equation}
  \Psi_i^{\text{C2HDM}} = (R_{i3})^2,
\end{equation}
where $R$ is the orthogonal rotation matrix that diagonalises the neutral mass-mixing matrix. The maximum amount of CP admixture observed for the SM-like scalar in the C2HDM is $\sim 9.5\%$ for $\xi_p > 1$ and $\sim 10\%$ for $\xi_p < 1$.

Similar to the C2HDM scenario, the distinctive resonance structures in the $M_{hh}$ and $M_{hhh}$ distributions then lead to enhancements of the triple Higgs production rate $\sim 20$ compared to di-Higgs production in the N2HDM (Fig.~\ref{ref:n2hdm_scans}). Most notably, some of these parameter points do not show overly anomalous behaviour in the $hh$ modes. This can be attributed to the fact that our scan is statistically limited by the number of points with the heavier exotics having a mass close to the $2 m_h$ threshold, where we observe the most enhancements in double Higgs production. Factors of $\lesssim 10$ enhancements in triple Higgs will be extremely challenging at the HL-LHC but represent an opportunity at, e.g., the FCC-hh. To this end, we note that all of our findings directly generalise to a potential FCC-hh, with reference cross sections of $\sigma_{hh}^{\text{SM}} \approx 800 ~\text{fb}$~\cite{FCC:2019hfo} and $\sigma_{hhh}^{\text{SM}} \approx 2.9 ~\text{fb}$~\cite{Papaefstathiou:2019ofh}.

Our findings are based on a leading-order exploration of the electroweak sector. Whilst QCD corrections are known to be relatively insensitive to the mass scales in multi-Higgs production (which is, in fact, exploited in most precision calculations there), weak radiative corrections might sculpt our findings quantitatively, see \cite{Kanemura:2016lkz,Braathen:2019zoh,Bahl:2023eau,Heinemeyer:2024hxa} for the impact on the 125 GeV self-coupling. These modifications are only part of the full (resonant) amplitude and warrant further investigation. Our observation related to the lightness of the degrees of freedom should be relatively robust against higher-order corrections.

\section{Conclusions}
\label{sec:conc}
The search for anomalies in multi-Higgs final states at the LHC is a theoretically motivated avenue to inform the beyond the Standard Model physics programme in the near future. New contributions to the effective Higgs potential have direct consequences for the electroweak history of the Universe: If strong enough, the electroweak phase transition could be part of an elegant explanation for the observed matter anti-matter asymmetry through electroweak baryogenesis. To leave a notable imprint in the thermal history of the Universe, exotic states should be relevant in the vicinity of the electroweak scale. Extra scalar degrees of freedom in this mass range as predicted by Higgs sector extensions (which can also include new sources of CP violation) could be experimentally opaque due to accidental signal background interference in top final states~\cite{Gaemers:1984sj,Dicus:1994bm,Bernreuther:1998qv,Jung:2015gta,Djouadi:2019cbm,Basler:2019nas}. These final states are typically the preferred decay modes of such scenarios~\cite{Basler:2018dac}. We take this as motivation to survey multi-Higgs production in the 2HDM and motivated extensions for parameter choices that lead to a strong first-order electroweak phase transition (see also the recent~\cite{Papaefstathiou:2023uum,Brigljevic:2024vuv}). Whilst di-Higgs production provides the phenomenologically most relevant avenue to detect such extensions experimentally at the high-luminosity phase, we highlight the relevance of triple-Higgs production, which can receive significant enhancement over the (phenomenologically irrelevant) SM rate at the LHC. In particular, in the singlet-extended 2HDM, the triple Higgs production rate can be around 40 times larger than the SM expectation, whilst only showing a modest and perhaps experimentally unresolvable deviation from the SM expectation in the $hh$ channels. The sensitivity range of the HL-LHC to triple Higgs production is yet to be analysed in realistic experimental projections. This will be undoubtedly challenging, but the level of enhancement that we observe in our parameter scan serves as a theoretical motivation to further pursue these efforts that are currently at an early stage of exploration by the ATLAS and CMS collaborations.

The observed enhancements are combinations of resonance structures in the Higgs final state kinematics: $H\to hh$ resonances are probed in a wider kinematic region in triple Higgs as compared to di-Higgs production. Furthermore, three-body decays $H\to hhh$ have the potential to dramatically increase the $hhh$ cross sections, albeit starting from an elusively small SM expectation. The biggest enhancements we observe in our scan combine these sources of $hhh$ cross-section enlargement. 

Whereas the deviations from the SM expectation in $hh(h)$ production are a consequence of kinematics, the lightness of the exotic spectrum parametrically controls the strength of the phase transition (in this work during the percolation stage). The correlation of multi-Higgs rates with the strength of the EWPT is therefore not a direct one but a light exotics mass scale has observable consequences for the multi-Higgs rates through new resonant cross section contributions. Additional degrees of freedom widen the 2HDM-expected correlations of couplings of the light (and SFOEWPT-relevant) scalar degrees of freedom. The correlated decay phenomenology of the heavy Higgs partners can result in significant enhancements of the multi-Higgs boson rates, in particular for $pp\to hhh$. Triple Higgs searches are undoubtedly challenging endeavours at the LHC, yet the typical enhancements that we can observe in extended Higgs sectors raise the cross section to regions where events will be recorded at the LHC, in particular with experimentally exploitable resonance structures from the Higgs decay kinematics. The futile outlook that might be based on the small non-resonant SM rate therefore appears to be overly pessimistic.

\medskip 
\noindent {\bf{Acknowledgements}} ---
We thank Apostolos Pilaftsis and Gilberto Tetlalmatzi-Xolocotzi for helpful discussions. C.E. also thanks the CERN Theory Department for their hospitality during the completion of this work.

This work is funded by a Leverhulme Trust Research Project Grant RPG-2021-031. L.B.~is partly supported by the BMBF-Project 05H21VKCCA. C.B.~and M.M.~acknowledge partial support by the Deutsche Forschungsgemeinschaft (DFG, German Research Foundation) under grant 396021762 - TRR 257. C.E. is supported by the UK Science and Technology Facilities Council (STFC) under grant ST/X000605/1 and the Institute for Particle Physics Phenomenology Associateship Scheme.  W.N. is funded by the University of Glasgow, CoSE Scholarship.

\bibliographystyle{apsrev4-2}
\bibliography{references}

\end{document}